\documentclass[10pt,aps,nofootinbib,superscriptaddress,twocolumn,preprintnumbers,balancelastpage]{revtex4-1}
\usepackage{amsmath,amssymb,amsthm,amsopn}
\usepackage{epsfig}
\usepackage{psfrag}
\usepackage[colorlinks=true,allcolors=blue!65]{hyperref}
\usepackage{multirow}
\usepackage{empheq}
\usepackage{slashed}
\usepackage{graphicx}
\usepackage{url}
\usepackage{subfigure}
\usepackage{textcomp}
\usepackage{bm}
\usepackage{dcolumn}
\usepackage{color,xcolor}
\usepackage{ulem}
\usepackage{cancel}
\usepackage{braket}
\usepackage[title]{appendix}
\usepackage{lipsum}
\usepackage{mathptmx}

\DeclareMathAlphabet{\mathcal}{OMS}{cmsy}{m}{n}

\def\blue#1{\textcolor{blue}{#1}}

\def\nn{\nonumber}

\def\comment#1{}

\def\beq{\begin{equation}}
	\def\eeq{\end{equation}}
\def\bea{\begin{eqnarray}}
	\def\eea{\end{eqnarray}}

\begin{document}
	\renewcommand{\topfraction}{1.0}
	\renewcommand{\bottomfraction}{1.0}
	\renewcommand{\textfraction}{0.0}
	
	\newcommand\sect[1]{\emph{#1}---}

	\title{Holographic complex potential of a quarkonium from deep learning }
	\author{Mahdi Mansouri}
	\email{mahdimansouri7577@gmail.com}
	\affiliation{Faculty of Physics, Shahrood University of Technology, P.O.Box 3619995161 Shahrood, Iran}
	
	\author{Kazem  Bitaghsir  Fadafan}
	\email{bitaghsir@shahroodut.ac.ir}
	\affiliation{Faculty of Physics, Shahrood University of Technology, P.O.Box 3619995161 Shahrood, Iran}
	
	\author{Xun Chen}
	\email{chenxunhep@qq.com}
	\affiliation{School of Nuclear Science and Technology, University of South China, Hengyang 421001, China}

	\begin{abstract}
	Utilizing an emergent metric developed from deep learning techniques, we determine the complex potential associated with static quarkonium. This study explores the disintegration process of quarkonium by analyzing the real component of this potential, which is crucial for understanding its stability in various conditions. We show that the dissociation length, the critical distance at which a quark and antiquark pair disintegrate, decreases as the temperature increases. Furthermore, our assessment of the imaginary component of the potential indicates an increase in the magnitude of the imaginary potential for quarkonium as temperatures rise. This enhancement contributes to the quarkonium's suppression within the quark-gluon plasma, mirroring the anticipated outcomes from QCD. Our findings not only confirm the theoretical predictions but also demonstrate the efficacy of deep learning methods in advancing our understanding of high-energy particle physics.
				
	\end{abstract}
	
	\maketitle

	\section{Introduction}
Recently, the application of machine learning to theoretical and experimental physics has been growing \cite{Carleo:2019ptp}. It would be a very interesting research field to use it in the context of the AdS/CFT correspondence \cite{Maldacena:1997re,Gubser:1998bc,Witten:1998qj} to study the dual bulk gravity. This can be achieved by considering the observables in the boundary theory as input data. Hashimoto et al. proposed an interesting deep learning (DL) and neural network approach to construct the bulk metric in this way \cite{Hashimoto:2018ftp,Hashimoto:2018bnb,Hashimoto:2019bih}. They were able to study the dynamics of a scalar field in a discretized curved spacetime as the architecture of a neural network. In this approach, the bulk gravity metric is represented by the network weights, and the input data are the VEV and the conjugate source, which come from the lattice data.  The application of the model has rapidly advanced, yielding intriguing results. These include extracting novel features from lattice QCD data of the chiral condensate, predicting the mass spectrum of excited state mesons, and deriving a dilaton potential, among others \cite{Hashimoto:2018bnb,Hashimoto:2019bih,Hashimoto:2020jug,Hashimoto:2020mrx}. The simplest examples has been explored in \cite{Song:2020agw} for someone who has not a background knowledge of holography. Different approaches have been studied, recently. For example, holographic flavor dependent critical endpoint of the QCD phase diagram has been studied through machine learning in \cite{Chen:2024mmd}. One also finds machine learning study of holographic black hole from lattice QCD equation of state in \cite{Chen:2024ckb}. The reconstruction of the bulk geometry from holographic conductivity and shear viscosity  have been done in \cite{Li:2022zjc} and \cite{Yan:2020wcd}, respectively.

This approach is termed the AdS/DL correspondence, which utilizes the lattice data of the chiral condensate to study holographic QCD setups. Generally, it introduces a novel method by implementing the AdS/CFT dictionary within a deep neural network. Such a network is considered as the dynamics of the field on a discretized curved background. The holographic radial direction in AdS/DL corresponds to the depth dimension of the network. The bulk spacetime is then viewed as the neural network, where the input data comes from the boundary field theory, and the weights in the deeper layers of the network are optimized. This process ultimately leads to the optimal bulk gravity for the boundary field theory. It is important to note that this is an inverse problem because, in the applied AdS/CFT, we first consider a gravity background and then solve it to find a geometry. We can then probe the background geometry to compute quantities that are dual to the observables in QCD.

In this paper, we utilize the emergent metric from AdS/DL to study the complex potential of a quarkonium. It is a bound state of a heavy quark and antiquark like charmonium $c\bar{c}$ or botomonium $b\bar{b}$. The physics of quarkoniums gives us a unique tool to simplify the complex physics of the strong interaction in QCD. Since instead of considering a quantum field theory description, one may study a Schrodinger equation with an effective complex potential to describe the physics of them at zero or non zero temperature \cite{Brambilla:2004jw}.

The motivation to study the holographic complex potential at finite temperatures stems from the melting of heavy quarkoniums like $ J/\psi $ and excited states in the quark-gluon plasma (QGP). It is found that their suppression is a significant experimental signature indicating whether QGP has been produced in heavy ion collisions \cite{Melting}. The primary mechanism proposed in \cite{Melting} responsible for this suppression is color screening. However, recent studies suggest a more significant factor than screening: the presence of an imaginary component in the heavy quark potential \cite{Burnier:2015tda}. Consequently, the thermal width of these systems has become a crucial topic in the study of QGP. In medium effects need separation scales between the rest mass of the quarks and environmental scales to improve the theoretical methods. The effective  field theory approaches are non-relativistic QCD or using potential non-relativistic QCD, see \cite{Brambilla:2004jw} for a review. Within this framework, thermal decay widths have been extensively analyzed. It has been shown that, at the leading order, two distinct mechanisms contribute: Landau damping and singlet-to-octet thermal breakup .

As a concrete example of the emergent metric's predictions from AdS/DL, the real part of the complex quark-antiquark potential has been examined in \cite{Hashimoto:2018ftp}. It has been also demonstrated that the real part of the heavy quark potential includes a linear confining term and the Debye screening part. It is important to notice that determining the imaginary part of the potential using lattice QCD presents more technical challenges compared to its real counterpart. However, research has established the existence of an imaginary component. Despite considerable uncertainties, the findings are consistent with theoretical model expectations \cite{Larsen:2024wgw}. The complex in medium potential has been also studied in Lattice \cite{Debnath:2023dhs}. Imaginary part of the potential exhibits non zero value above the deconfined temperature. By using the Schrodinger equation with complex potentials obtained from lattice QCD, the static properties and dynamical evolution of bottomonium states in the QGP has been studied in \cite{Chen:2024iil}. These approaches allow for a nuanced understanding of how quarkonium states behave and evolve under the extreme conditions present in the QGP.

Within the framework of AdS/CFT, some methodologies can result in a complex static potential. For example, the complex heavy quark potential has been revisited in \cite{Albacete:2008dz}. It has been found that there is a kink above a certain critical separation. Interestingly, it is shown that by employing a different renormalization subtraction and by analytically continuing the U-shaped string into the complex plane, a smooth, non-zero heavy quark potential can be achieved without such a kink. As a significant observation, the derived heavy quark potential also possesses a non-zero imaginary part. For our analysis, the source of the imaginary component of the potential is the fluctuations occurring at the lowest point of the U-shaped configuration of the string within the bulk\cite{Noronha:2009da}. Employing this technique, we study the imaginary potential from emergent metric.

Holographic study of the imaginary potential and quarkonium physics have been done in \cite{Noronha:2009da,Finazzo:2013rqy}. The impact of the potential's imaginary component on the thermal widths of states in both isotropic and anisotropic plasmas has also been investigated from holography in \cite{BitaghsirFadafan:2013vrf}. The effects of charge and finite 't Hooft coupling correction on the imaginary potential of a static and moving heavy quarkonium have been studied in \cite{Fadafan:2013coa} and \cite{BitaghsirFadafan:2015yng}, respectively. Recently, bottomonia supperation in a strongly coupled QGP has been computed for the first time in \cite{Barnard:2017tld} and the ground state bind energy computed from holography. In a rotating matter, holographic imaginary potential of heavy quarkonia was studied in \cite{Zhang:2023psy}. One finds effect of gluon condensation in \cite{Tahery:2022pzn}.

The use of DL to understand the real and the imaginary part of the heavy quarkonium potential is an interesting area of research \cite{Shi:2021qri}. Deep learning models, particularly neural networks, can be trained to recognize patterns and make predictions based on large datasets. In the context of heavy quarkonium, these models could potentially learn from data generated by lattice QCD simulations or from experimental results to predict the behavior of the imaginary potential. By comparing the results obtained from deep learning models with those from traditional field theory approaches, the new models can be checked and understanding of QGP dynamics improved. It also might help in understanding non-perturbative effects that are difficult to calculate using standard field theory methods. It's important to note that while deep learning offers powerful tools for analysis and prediction, the results need to be interpreted with caution. The models are only as good as the data they are trained on, and they require proper validation against known physical laws and experimental data. Moreover, the interpretability of DL models is often a challenge, making it essential to use these tools in conjunction with traditional analytical methods to gain a comprehensive understanding of the phenomena. Thus, it gives us a good motivation to consider emergent metric developed from AdS/DL and study if the imaginary part of the heavy potential reproduces the expected results from QCD. 

The paper is organized as follows, we review the emergent spacetime from AdS/DL duality in section \eqref{sectionI}. The real and imaginary parts of the quarkonium will be coaculated in section \eqref{sectionII}. We summerize the discussions and conclusions in section \eqref{sectionIII}.

\section{Review of emergent metric from AdS/DL  }\label{sectionI}
In this section we review the proposed neural network in \cite{Hashimoto:2020jug}. First we introduce the background metric. The bulk spacetime coordinates are $4+1$ dimensions denoted by $(t, \eta, x_1,x_2,x_{3})$ where the holographic radial dimension is $\eta$. The boundary field theory has a translational symmetry and the background metric in the string frame is given by the following equation

\begin{equation}
\mathrm{d}s^{2} = -f(\eta)\mathrm{d}t^{2}+\mathrm{d}\eta^{2}+g(\eta)(\mathrm{d}x^{2}_{1}+\mathrm{d}x^{2}_{2}+\mathrm{d}x^{2}_{3}) \, .
\label{metric}
\end{equation}
The black hole horizon is located at $\eta=0$ and the boundary field theory at $\eta=\infty$. We consider the radial direction $\eta$ in the range $ 0.15 < \eta < 1$.
\begin{figure*}[!ht]
	{\centering%
		\begin{tabular}{@{}cc@{}}
			\includegraphics[width=70mm]{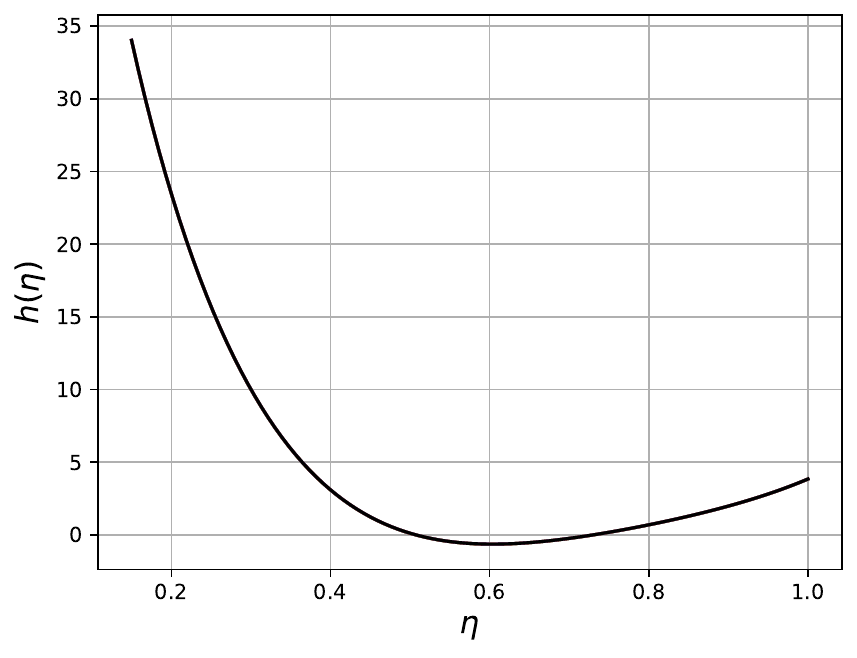}&
			\includegraphics[width=70mm]{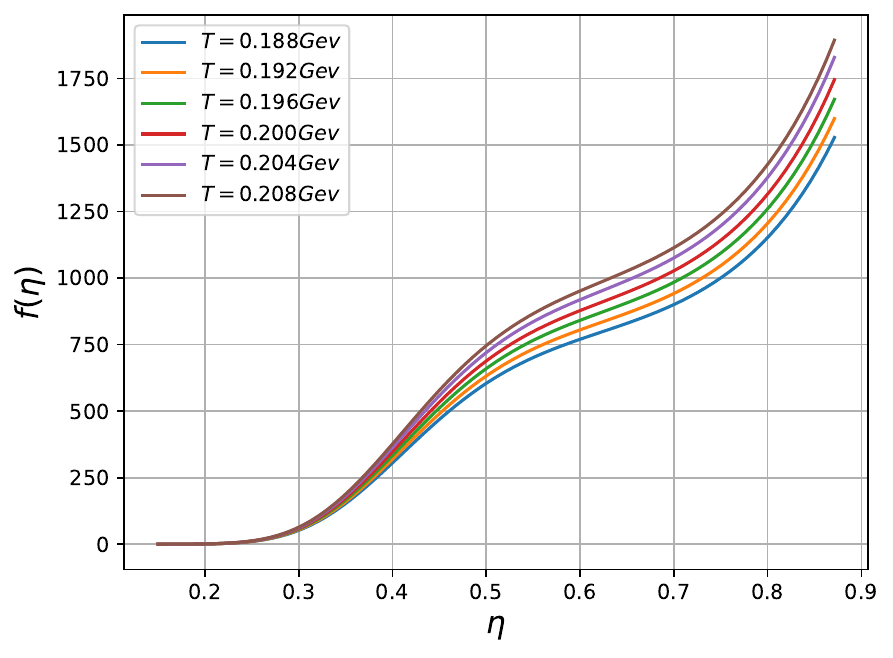}
		\end{tabular}\par}
	\caption{The emergent deep learning metric functions as a function of the radial direction. $h(\eta)$ is independent of temperature but $f(r)$ changes at different temperatures. }
	\label{fighf}
\end{figure*} 
A scalar field $\phi$ is introduced in the background which is described by the following action with $\phi^4$ interaction as
\begin{equation}
\begin{split}
S[\phi] = \dfrac{1}{2}\int d\eta \sqrt{f(\eta)g(\eta)^3}\left( g^{\mu \nu}\partial_{\mu}\phi \partial_{\nu}\phi+m^{2}\phi^2+\frac{\lambda}{2}\phi^4\right) \, .
\end{split}
\end{equation}
The equations of motion of $\phi$ read
\begin{equation}
\begin{aligned}
&\pi = \partial_{\eta}\phi \, ,\\
&\partial_{\eta}\pi+\pi\, \partial_{\eta}\ln\sqrt{f(\eta)g(\eta)^{3}} -m^{2}\phi -\lambda \phi^3 = 0 \, .\label{EOM}
\end{aligned}
\end{equation}
For simplicity of doing numeric, a new combination of the metric functions is defined as
\begin{equation}
h(\eta) \equiv \partial_{\eta}\ln\sqrt{f(\eta)g(\eta)^{3}} \, .
\label{hdef}
\end{equation}
The input data is given by $\phi(\infty)$ and $\pi(\infty)$ at the boundary of the AdS spacetime. One expects at horizon $\pi(0)=0$. Based on the AdS/CFT dictionary, the boundary behavior of the scalar field $\phi$ with the mass $m^2=\frac{-3}{L}$ gives us the condensate in the presence of the source. 

The positive data is given by the experimental pair of data at the boundary as $(\phi(\infty),\pi(\infty))$. One should notice that they satisfy the black hole boundary condition and the dynamics equations in \eqref{EOM}. But one needs also the negative data which are those data that do not obey black hole boundary condition and do not lie on the experimental curve.  Thus, for a given pair of experimental data, there are three parameters to be learned as 
\begin{itemize}
	\item The AdS radius $L$
	\item Interaction coupling $\lambda$
	\item The metric function $h(\eta)$
\end{itemize}
The neural network is equations of motion in \eqref{EOM}, the weights in the network are $h(\eta)$. The training data are the lattice data in which gives us the chiral condensate as a function of the quark mass at fixed temperature. The chiral condensate $\langle\bar{\psi}\psi\rangle$ is given in units of $[({\rm GeV})^{3}]$ and quark mass $m_q$ is in [GeV]. Fixed temperatures are $T=0.188$, $0.192$, $0.196$, $0.200$, $0.204$ [GeV], which is converted to physical units \cite{Hashimoto:2018bnb}. 

The Neural Ordinary Differential Equation (ODE) is the framework for finding the optimized estimation for $h(\eta)$, $L$ and $\lambda$. \footnote{To review Neural ODE see \cite{Hashimoto:2020jug}. } In \cite{Hashimoto:2022eij}, the dilaton potential has been derived from the above machine learning method. First they find the emergent bulk spacetime as an inverse problem from the QCD lattice data as the boundary data. Then move to the next inverse problem which is finding a bulk gravity action whose solution is given by the emergent metric function.

\subsection{Emergent metric }
The emergent metric functions in \eqref{metric} were found by the lattice QCD data of the chiral condensate using the neural network ODE in \cite{Hashimoto:2020jug}. After training, it is found that $L=5.164$ but $\lambda$ depends on the temperature as Table \ref{dataL}.
\begin{table}[ht]

	\centering
	\begin{tabular}{|c||c|c|c|c|c|c|}
		
		\hline
		\centering
		\small
		\setlength\tabcolsep{3pt}
		\blue{$T$} &\blue{ 0.188} &\blue{ 0.192} & \blue{0.196} & \blue{0.200} & \blue{0.204} & \blue{0.208}\\ 
		\hline
		$\lambda$ & 0.0014 & 0.0011 & 0.0009 & 0.0007 & 0.0005 & 0.0003\\ \hline
	\end{tabular}	\caption{Deep learning results of the interaction coupling $\lambda$ at different temperatures. One finds AdS radius $L=5.164~ GeV^{-1}$.}
	\label{dataL}
\end{table}

Interestingly, at low temperature the system is strongly coupled which is related to the self coupling of sigma meson \cite{Hashimoto:2020jug}. Also  it is found that $h(\eta)$ does not depend on the temperature which is not surprising because it is given in terms of $f(r)$ and $g(r)$ in \eqref{hdef}. We will see that  $f(r)$ and $g(r)$ depend on the temperature. The trained metric function $h(\eta)$ is found to have the following expression
\begin{align}
	&h({\eta}) =2.84 -24.14\,(1-\eta) + 55.62\,(1-\eta)^2 - \nonumber\\&130.22\,(1-\eta^3) +150.79 \,(1-\eta)^4
	+5.57\,(1-\eta)^5 +\nonumber\\& 4.68\,(1-\eta)^6 +  3.56\,(1-\eta)^7 +2.53\,(1-\eta)^8 +\frac{1}{\eta}.
	\label{secondc}
\end{align}
The last term is needed because of the divergency of the metric function at the horizon $\eta=0$. As it was expressed the horizon is located at $\eta=0$ but the radial direction $\eta$ in \cite{Hashimoto:2020jug} changes in the the range $0.1 < \eta < 1 $. It is technically difficult to study $\eta=0$ behavior. To better approximate this region, this range was extended in \cite{Hashimoto:2022eij} as $0.01 < \eta < 1 $. In this paper we choose the minimum value of the range of $\eta=0.15$. The ansatz for $h(\eta)$ in \cite{Hashimoto:2022eij} is not the same as \cite{Hashimoto:2020jug} because they study the Einstein dilaton gravity in which only the odd powers are allowed for $h(\eta)$. It was mentioned in \cite{Hashimoto:2022eij} that the trained bulk spacetime is quite similar to one obtained in \cite{Hashimoto:2020jug}. In this paper we follow the approaches in \cite{Hashimoto:2020jug}.

One needs $f(\eta)$ and $g(\eta)$ to calculate the Wilson lines from the AdS/CFT correspondence. In \cite{Hashimoto:2020mrx}, it was shown how we can derive the bulk metric functions from Wilson loops. The metric function $g(\eta)$ is chosen to be similar to the AdS Schwarzschild solution as 
\begin{align}
	g(\eta) = A \left(2\cosh\frac{2\eta}{La}\right)^a \, ,
	\label{gansatz}
\end{align}
where $A=\left( 2 \pi T L \right)^2 e^{c(a(T))}$. The quantity $c(a(T))$ is found by solving $\frac{\partial}{\partial T} \left( T^2  e^{c(a(T))} \right)=0 $ numerically where $a$ is a temperature-dependent constant \cite{Hashimoto:2020jug}. 

The metric function $f(\eta)$ is given by the following expression
\begin{align}
	f(\eta) &= \left( 2\pi T L\, tanh (\eta/L)\right)^2\,\nonumber \\&
	exp \int _0^\eta 2\left( h(\eta)- \frac{2}{L\,sinh(2\eta/L)} -\frac{3}{L}tanh2\eta/La \right)d\eta  \, .
	\label{fdef}
\end{align}
Given the metric functions in \eqref{gansatz} and \eqref{fdef}, one can study the Wilson loop calculations in the next section.

The metric function $h(\eta)$ and $f(\eta)$ have been plotted in Fig. \ref{fighf}. While $h(\eta)$ does not depend on the temperature, the metric function $f(\eta) $ changes smoothly at different temperatures.

\section{The Complex Heavy quark potential} \label{sectionII}
Given the emergent bulk spacetime metric in the previous section, one can study some predictions via the AdS/CFT dictionary. For exampple, using this approach and from expectation values of Wilson loops, one finds a bulk IR bottom for any confining heavy quark potential \cite{Hashimoto:2020mrx}. Here we present the results for the complex quark antiquark potential. The real part of the potential was already studied in \cite{Hashimoto:2020jug}, first we study its general feature and next we compute the imaginary part.

\subsection{Real Part of $	V_{Q\bar{Q} }$ }
We consider the quark and antiquark on the boundary of the emergent AdS spacetime. The separation distance between them is $d$ and a U-shaped string connects them to each other. The deepest point of the string in the bulk is $\eta=\eta_0$. Based on  \cite{Maldacena:1998im,Rey:1998ik,Kinar:1998vq}, the distance $d$ as a function of $\eta$ is given by 
\begin{align}
	d = 2\int_{\eta_0}^\infty
	\frac{1}{\sqrt{g(\eta)}}\sqrt{\frac{f(\eta_0)g(\eta_0)}{f(\eta)g(\eta)-f(\eta_0)g(\eta_0)}}d\eta \, ,
	\label{d}
\end{align}
\begin{figure}[!ht]
	{\centering%
		\begin{tabular}{@{}cc@{}}
			\includegraphics[width=70mm]{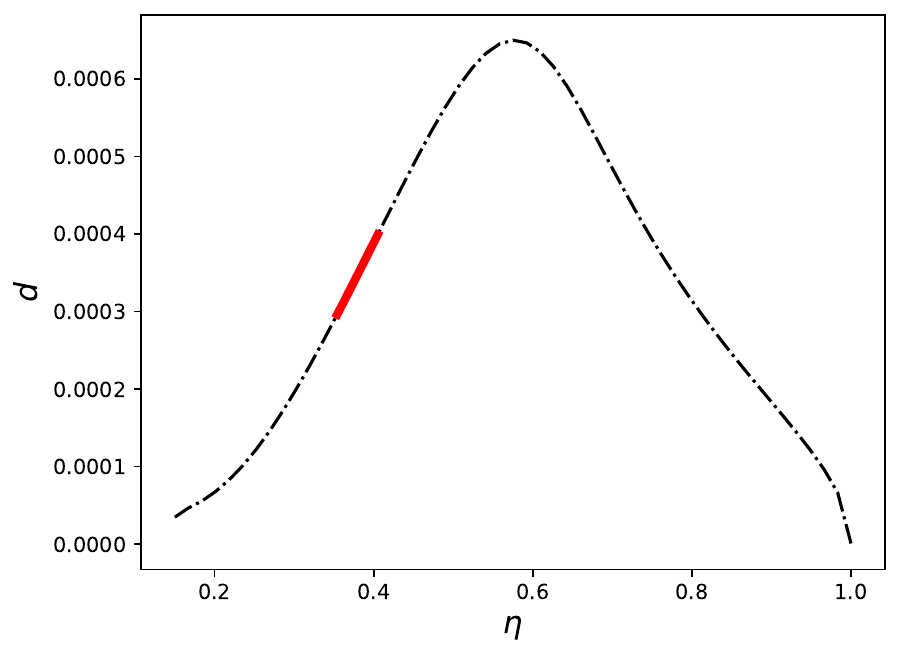}&
			\end{tabular}\par}
	\caption{The quark antiquark distance as a function of the radial coordinate $\eta$ at $T=0.188 GeV$. The red color part shows the distance that generates the imaginary part of the potential. }
	\label{deta}
\end{figure}
Using the trained metric from AdS/DL, the explicit form of $d$ is calculated in Figure \ref{deta}. It shows the quark and antiquark distance $d$ in the units of Fermi as a function of the radial direction $\eta$. From this figure, one finds the expected results that originates from holography, there is a maximum distance for the quark antiquark separation so that the connecting U-shape string breaks for $d$ larger than this maximum distance to two straight strings. Also for distance smaller than this maximum value, there are two U-shape solutions. One is called short string and the other one long string. Only the short strings are stable against the fluctuations on the connected strings. 

\begin{figure}[!ht]
	{\centering%
		\begin{tabular}{@{}cc@{}}
			\includegraphics[width=70mm]{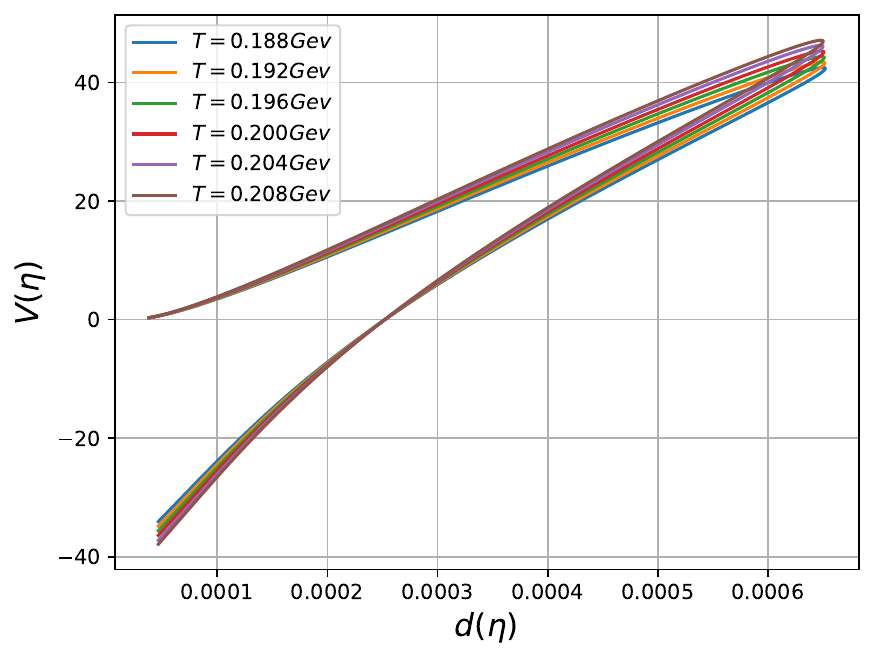}
		\end{tabular}\par}
	\caption{The real part of heavy quark antiquark potential versus the radial coordinate $\eta$ at different temperatures.}
	\label{Vdd}
\end{figure}

The real part of the heavy quark potential is given by the following equation \cite{Maldacena:1998im,Rey:1998ik,Kinar:1998vq}
\begin{align}
2\pi \alpha'\,Re ~V_{Q\bar{Q} }&=& 2\int_{\eta_0}^\infty \!\!\!\!
	\sqrt{f(\eta)}\left(
	\sqrt{\frac{f(\eta)g(\eta)}{f(\eta)g(\eta)-f(\eta_0)g(\eta_0)}} \right)d\eta \nn\\&&
	-2\int_{0}^\infty \sqrt{f(\eta)}d\eta \, .
	\label{Vd}
	\
\end{align}
In Figure \ref{Vdd}, we abbreviate $ 2 \pi \alpha' Re~V_{Q\bar{Q} }$ as $V$ and plot it  as a function of $d$ at different temperatures. At zero temperature, the heavy quark potential is of Coulomb type due to the conformal invariance of the theory. However, at finite temperatures, the behavior changes notably. The potential starts out similar to the zero-temperature case for small separations between the quark and antiquark but rises steeper than the vacuum potential, eventually becoming zero at a certain critical separation distance. Beyond this distance, the potential was initially thought to vanish, suggesting a "melting" of the string due to the presence of a black hole horizon in the AdS space. As mentioned, recent studies have revisited this calculation, pointing out that by analytically continuing the string configurations into the complex plane and using a different renormalization subtraction, one can obtain a smooth, non-zero potential that does not exhibit a kink at the critical separation \cite{Albacete:2008dz}. 

\begin{figure}
	\centering
	\includegraphics[width = 0.8\linewidth]{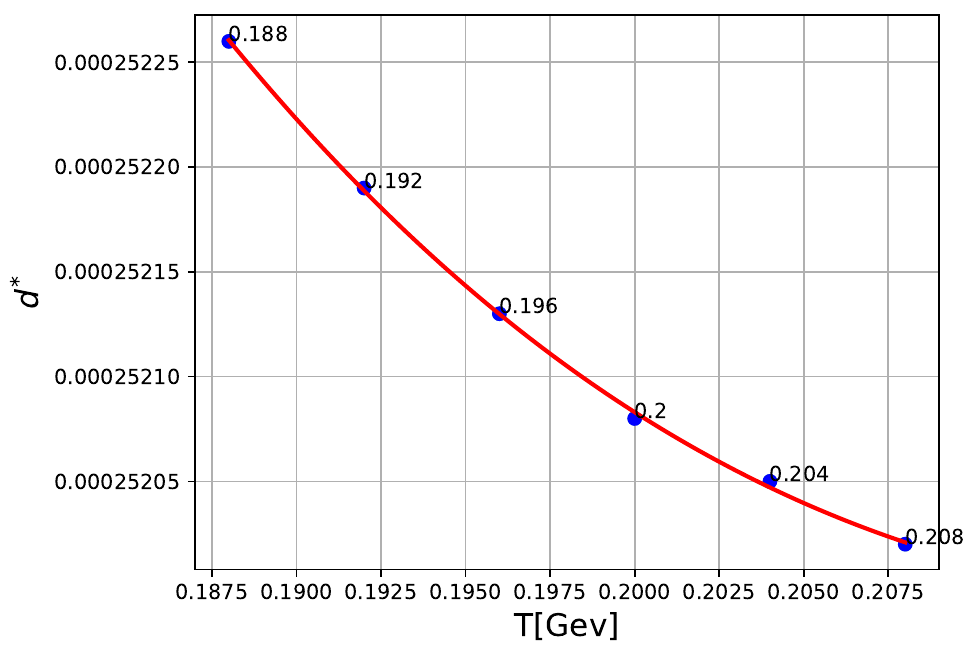}
	\caption{The dissociation length versus the temperature.}
	\label{dstar}
\end{figure}

In \cite{Ali-Akbari:2009svi}, melting of heavy mesons have been studied. It was found that by increasing temperature mesons melt sooner. From Figure \ref{Vdd}, meson melting occurs at $d=d_*$ when $Re~V_{Q\bar{Q} }$ vanishes. This distance is called dissociation length and we plot it versus the temperature in Figure \ref{dstar}. As anticipated from quarkonium physics, an increase in temperature leads to the dissolution of quarkonium states, particularly those with smaller interquark distances. From perturbative QCD, at high temperatures, color screening in the quark-gluon plasma reduces the effective interaction between quarks color screening leading to the weakening and eventual dissolution of quarkonium states \cite{Burnier:2015tda}.

\subsection{Imaginary Part of $	V_{Q\bar{Q} }$ }
We have analyzed the imaginary potential and the thermal width of static and moving quarkonium through the AdS/CFT correspondence in \cite{Fadafan:2013coa,BitaghsirFadafan:2015yng}. According to the framework of \cite{Noronha:2009da}, the thermal width of heavy quarkonium states arises from thermal fluctuations, which are a consequence of the interactions between heavy quarks and the strongly coupled medium. This methodology was further examined in \cite{Finazzo:2013rqy}, where general conditions for the emergence of an imaginary component in the heavy quark complex potential were established. Alternative methods have been proposed in \cite{Albacete:2008dz,Hayata:2012rw}. There are some limitations to the calculations of the holographic imaginary potential and the	thermal width in this method \cite{Finazzo:2013rqy,BitaghsirFadafan:2013vrf}. The scenario involving anisotropic plasma has been scrutinized in \cite{BitaghsirFadafan:2013vrf}, leading to the derivation of an imaginary potential formula applicable in a general curved background. Using it, one finds the imaginary potential in the emergent spacetime background as 

	\begin{align}\label{ImVdeff}
2\pi \alpha'\,	\text{Im} V_{Q\bar{Q}}=-\left(\frac{(f(\eta_{0})g(\eta_{0}))'}{2f(\eta_{0})g(\eta_{0}))''}-\frac{f(\eta_{0})g(\eta_{0})}{(f(\eta_{0})g(\eta_{0}))'}\right)\sqrt{f(\eta_{0})}
\end{align} 

\begin{figure}
	\includegraphics[width=70mm]{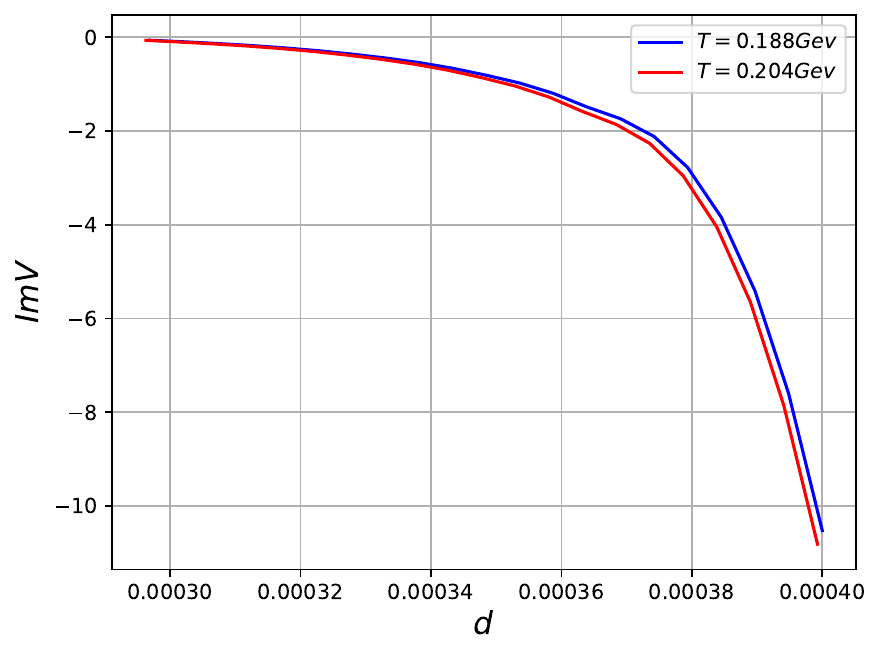}
	\caption{The imaginary part of the quark anti quark as a function of the radial coordinate $\eta$ at $T=0.188 GeV$ and $T=0.204 GeV$ from top to bottom.}
	\label{ImV-d}
\end{figure}

The onset of the imaginary part of the potential typically occurs at a certain distance, which is related to the medium's Debye screening length. In Figure \ref{deta}, we highlighted this distance by red color. In this range \eqref{ImVdeff} is negative as one expects from the imaginary potential definition. In the context of heavy quarkonium, it is negative due to its association with the dissipative processes in the QGP. It represents the thermal decay width of the quarkonium states, which is related to the probability of the quarkonium dissociating into unbound quarks within the plasma. In Figure \ref{ImV-d}, we plotted the imaginary potential as a function of the quark distance at two different temperatures, $T=0.188$ and $T=0.204$ GeV. 
In \cite{Chen:2024iil}, a large imaginary potential results in the suppression of bottomonium in the QGP. The width of the peak in the spectral function of heavy quarkonium can be translated into the imaginary part of the complex potential. The behavior of the imaginary potential of a quarkonium system at finite temperatures, as described in Figure \ref{ImV-d}, is consistent with the theoretical frameworks and computational approaches. In this figure, $2\pi \alpha'\,	\text{Im}$ abbreviated to $ImV$ in the vertical axis. Comparing with \cite{Larsen:2024wgw}, one finds that the imaginary potential increases with temperature, which is indicative of the destabilizing effects on quarkonium states in a hot medium. The increase in the absolute value of $ImV$ with temperature implies that thermal effects enhance the interaction energy between the quark and antiquark. This could be due to increased thermal agitation or the presence of additional degrees of freedom at higher temperatures, which contribute to the potential. Holographic models have also been found patterns similar to Figure \ref{ImV-d} as \cite{Finazzo:2013rqy}.

\section{Estimating the Thermal Width}
In this section, we estimate the thermal width of a heavy quarkonium, such as the $\Upsilon$ meson, using the emergent metric derived from machine learning. We follow a similar approach to the static case studied in \cite{Noronha:2009da,BitaghsirFadafan:2015yng}; however, we now incorporate the emergent metric into the calculation. 

The thermal width $\Gamma_{Q\bar{Q}}$ is determined by the imaginary part of the potential using a first-order non-relativistic expansion as \cite{Noronha:2009da}

\begin{equation}
	\label{eq:thermalwidth}
	\Gamma_{Q\bar{Q}} = - \langle \psi | \mathrm{Im} \, V_{Q\bar{Q}}(L,T) | \psi \rangle,
\end{equation}
where the ground state wave function in a Coulomb-like potential is given by
\begin{equation}
	\label{eq:wavefunction}
	 |\psi \rangle = \frac{1}{\sqrt{\pi} a_0^{3/2}} e^{-r/a_0}.
\end{equation}
Here, $a_0$ is the Bohr radius, and $m_Q$ is the mass of the heavy quark $Q$, such that $m_Q \gg T$. The thermal width is then given by
\begin{equation}
	\label{eq:thermalwidth2}
	\Gamma_{Q\bar{Q}} = -\frac{4}{a_0^3} \int_0^{\infty} dL \, L^2 e^{-2L/a_0} \, \mathrm{Im} \, V_{Q\bar{Q}}(L,T).
\end{equation}

When plotting Figure \ref{ImV-d}, we have taken into account the condition that the $\mathrm{Im} \, V_{Q\bar{Q}}(d)$ is negative, and used only the appropriate negative part of the imaginary potential. Although this is a simplification, the overall features of the curve in this figure qualitatively agree with the lattice calculations presented in \cite{Larsen:2024wgw}.

\begin{figure}
	\includegraphics[width=70mm]{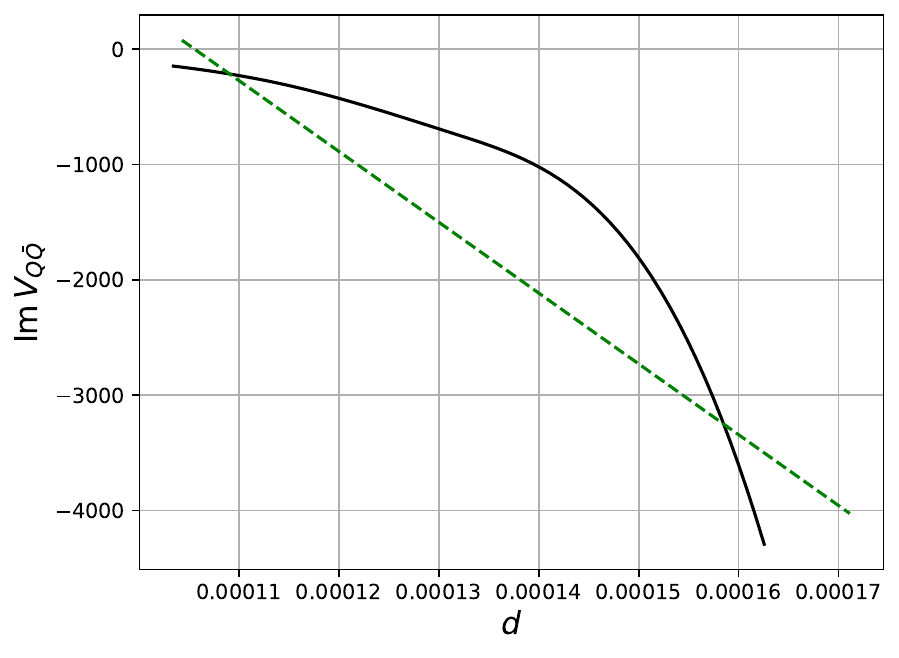}
	\caption{Fitted imaginary potential as a function of the quark-antiquark separation distance $d$. The dashed line corresponds to the straight-line fitting of the imaginary potential at $T = 0.208 \, \text{GeV}$.}
	\label{ImV-fitinm}
\end{figure}

Furthermore, as shown in Figure \ref{deta}, the imaginary potential exists only within a finite range of the distance between the quark and anti-quark, specifically between $d_{min}$ and $d_{max}$. This indicates that the imaginary potential is not present at all separations, but only within a specific interval. By focusing on the region between $d_{min}$ and $d_{max}$, one finds that there is a limitation in estimating the thermal width since we cannot take the integral in \eqref{eq:thermalwidth2} from zero to infinity. One may use a reasonable extrapolation, such as straight-line fitting or polynomial fitting, as shown in Figure \ref{ImV-fitinm}. For example, if we consider $T = 208 \, \text{MeV}$, using a fifth-order polynomial fitting, one gets the following relation:
\begin{eqnarray}
\mathrm{Im} \, V_{Q\bar{Q}}(d)& =& 2.23 \times 10^{21} \, d^{5} - 5.061 \times 10^{17} \, d^{4} \nonumber\\&&+ 5.705 \times 10^{13} \, d^{3} \allowbreak - 3.192 \times 10^{9} \, d^{2}\nonumber\\&& - 3.913 \times 10^{24} \, d + 7.093 \times 10^{4},
\end{eqnarray}

where $d$ is the quark-antiquark separation distance. 

\begin{figure}
\includegraphics[width=70mm]{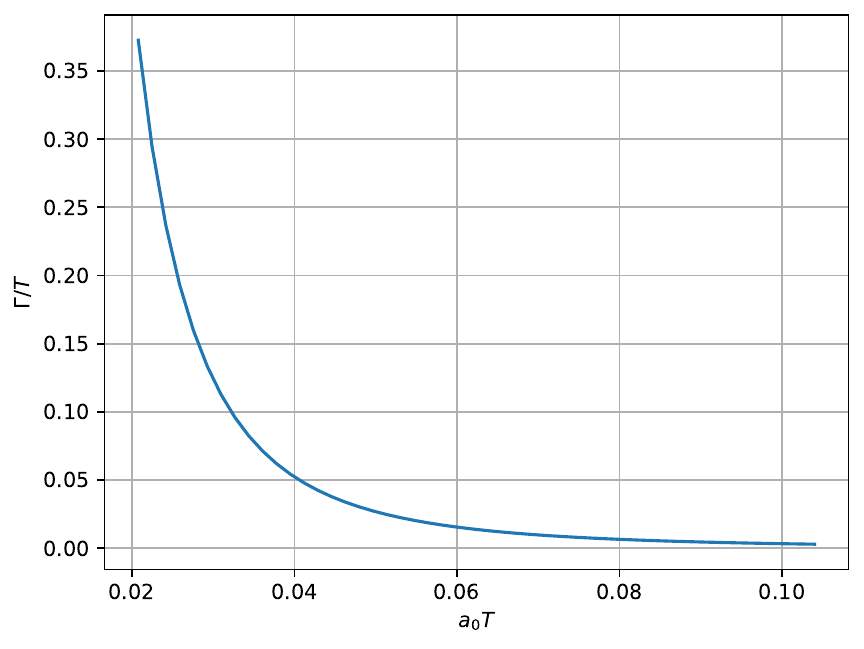}
\caption{The thermal width of the $\Upsilon$ divided by the temperature $T$ versus $T a_0$, where $a_0$ is the Bohr radius.}
\label{Thermalfig}
\end{figure}

On the other hand, in \eqref{eq:thermalwidth2}, the Bohr radius $a_0$ is a free parameter. One may try different values of $a_0$ to estimate $\Gamma_{Q\bar{Q}}$. We have shown its behavior as a function of the dimensionless $a_0 T$ in Figure \ref{Thermalfig}. One finds the same plot using straight-line fitting as well. For a reasonable value in the range of a few MeV for $\Upsilon$, we fix $a_0 \sim 0.6 \, \mathrm{GeV}^{-1}$ as in \cite{Finazzo:2013} and evaluate \eqref{eq:thermalwidth2} in the range of $L_{\mathrm{min}} = 0.000085$ and $L_{\mathrm{max}} = 0.00031$. In this case, the thermal width value is $0.532 \, \mathrm{MeV}$. Here, the temperature is $T=208 \, \text{MeV}$. 

\section{Summary}\label{sectionIII}
Quarkonium states are key probes for experimental studies of the QGP. Theoretical models of quarkonium production in high-energy nuclear collisions are heavily reliant on understanding the in-medium interactions between quark and antiquarks. These interactions are complex and are influenced by the QGP that is formed during such collisions. The potential that describes the interaction between the quark and antiquark is typically divided into real and imaginary components, each dependent on the temperature of the medium and the spatial separation of the quarks. The real part of the potential is associated with the binding force between the quarks, while the imaginary part is related to the dissociation processes and the thermal decay width of the quarkonium states. The behavior of these components potentials is crucial for predicting the yields and suppression patterns of quarkonium states in the QGP.

Using the emergent spacetime from AdS/DL, we studied the real and imaginary parts of the quarkonium potential. This method combines the concepts of holographic duality with the computational power of DL. It is known that this duality is particularly useful for studying strongly coupled systems, such as the QGP, which are difficult to analyze with perturbative methods. Recently deep neural networks have been applied to lattice quantum chromodynamics to obtain the temperature-dependent complex potential \cite{Shi:2021qri}. Such studies allow for a more accurate characterization of the in-medium interactions and enhances our theoretical understanding of quarkonium production in high-energy nuclear collisions.

We found that the quark antiquark distance, denoted as the dissociation length, decreases by increasing temperature. It implies that the quarkonium states needs more energy to dissociate at higher temperatures. We have also calculated the imaginary part of the potential and show that its absolute value increases by increasing the temperature. It was shown that the expected results from QCD can be derived from the holographic complex potential. In summary, the absolute value of the imaginary potential of a quarkonium increases with temperature, leading to greater suppression of it in the QGP. This behavior is crucial for using quarkonium suppression as an indicator of QGP formation and for estimating the temperature of the QGP in high-energy nuclear collisions. Also, we estimated the thermal width of a heavy quarkonium, such as the $\Upsilon$ meson, using the emergent metric. By incorporating the emergent metric into our calculations and exploring various values for the Bohr radius, we demonstrated the behavior of the thermal width under different conditions. Our findings qualitatively agree with existing literature, and we provided a reasonable estimation method for the thermal width within specific parameter ranges. 

Using the AdS/CFT correspondence, melting of quarkoniums and excited states have been studied in \cite{Fadafan:2012qy}. It was found that higher temperature needed to melt excited heavy quarkoniums. That would be interesting to use the trained metric by AdS/DL and investigate if such behavior can also be seen in this emergent background. Recently, based on the holographic model and using machine learning techniques which incorporates the equation of state and baryon number susceptibility for different flavors, the calculation of the drag force, jet quenching parameter, and diffusion coefficient of the heavy quark have been done in \cite{Chen:2024epd} at finite temperature. It is interesting to extend this study and consider imaginary potential of a moving quarkonium. Because when the quarkonium is moving, the valid region in Figure \ref{deta} can change. The velocity affects the screening of the potential and, consequently, the distance at which the imaginary part becomes significant. Such extensions would be interesting from phenomenological point of view and may be essential for theoretical predictions and comparison with lattice data. Our findings not only confirm the theoretical predictions but also demonstrate the efficacy of deep learning methods in advancing our understanding of high-energy particle physics.
\\
\section*{Acknowledgment}
This work is supported by the Natural Science Foundation of Hunan Province of China under Grants No. 2022JJ40344, the Research Foundation of Education Bureau of Hunan Province, China (Grant No. 21B0402).
	
\end{document}